\documentclass[british,aps,pra,reprint,longbibliography,floatfix]{revtex4-2}
\usepackage[T1]{fontenc}
\usepackage[utf8]{inputenc}
\usepackage{color}
\usepackage{babel}
\usepackage{amsmath}
\usepackage{amssymb}
\usepackage{graphicx}
\usepackage[bookmarks=false,
 breaklinks=false,pdfborder={0 0 1},backref=false,colorlinks=true]
 {hyperref}
\hypersetup{
 linkcolor=blue,citecolor=blue,urlcolor=blue}

\makeatletter

\usepackage{color}
\usepackage{babel}

\usepackage{bm}\usepackage{physics}\usepackage{tikz}
\usetikzlibrary{arrows.meta}

\def\@bibdataout@aps{%
 \immediate\write\@bibdataout{%
  @CONTROL{%
   apsrev42Control%
   \longbibliography@sw{%
    ,author="08",editor="1",pages="1",title="0",year="1"%
   }{%
    ,author="08",editor="1",pages="1",title="",year="1"%
   }%
  }%
 }%
 \if@filesw
  \immediate\write\@auxout{\string\citation{apsrev42Control}}%
 \fi
}%

\makeatother

\begin{document}
\title{Reaffirming a Challenge to Bohmian Mechanics}
\author{Jan Klaers}
\author{Violetta Sharoglazova}
\author{Marius Puplauskis}
\affiliation{Adaptive Quantum Optics, MESA$^{+}$ Institute of Technology, University
of Twente}
\begin{abstract}
In our work \citep{sh25}, we reported the first measurement of the
speed of tunnelling particles by comparing two orthogonal motions
in a coupled-waveguide system. Here we respond to objections raised
in comments, referee reports, preprints, and articles. In particular,
we distinguish two questions that are often conflated: whether Bohmian
mechanics reproduces the measured density, and whether the standard
guiding equation assigns the correct state of motion to the particles.
The first point follows under the usual quantum equilibrium assumptions.
By contrast, identifying the guiding velocity with particle motion
is an additional physical postulate, since the standard guiding equation
does not follow from the Schrödinger equation alone. We argue that,
in the evanescent regime, the state of motion assigned by the standard
guiding equation is in disagreement with the speed following from
our experiments. To make the distinction explicit, we also present
a bidirectional Bohmian model that reproduces the same stationary
density while assigning finite speeds compatible with the speed inferred
in the evanescent regime. 
\end{abstract}
\maketitle

\section{Purpose and scope}

This article collects and systematises our response to objections
that have been raised against the interpretation of our experiment
\citep{sh25}. The objections and related discussions have appeared
in comments, preprints, and articles \citep{Dr26,Beck26,Ni25,Sienicki25,Di25,Wa25,Dr25,Waegell26,YeMeasured25,YeActual26,ShangInstant25,DM26,Daem26,DrNabet26}.
Beyond these replies, our experiment has also entered broader discussions
of tunnelling, the measurement problem, Bohmian mechanics, weak measurements,
quantum-optical analogues, and gravitational physics \citep{Fedrizzi25,GibneyTunnel25,GibneySurvey25,Tomaz25,Lantigua25,Mousavi25,Dona25,Seren25,Grande26,Shang26,XuWeakValues26,Danielsson26}.
The present article focuses on the former group of papers.

The debate surrounding our work often conflates two distinct questions.
The first is whether Bohmian mechanics can reproduce the measured
particle density. Under the usual quantum equilibrium assumptions,
it can. The second is whether the standard guiding equation assigns
the correct state of motion to the particles. Our experiment addresses
the second question. In the evanescent regime, the outcome of a physically
plausible speed measurement does not agree with the motional state
assigned by the standard guiding equation. The issue is therefore
not a failure of density reproduction, nor a rejection of trajectory-based
interpretations in general, but the physical role of the standard
guiding equation as a law of motion.

With this scope in mind, Section \ref{sec:background} sets the stage
by discussing how particle motion is represented in standard quantum
mechanics and in Bohmian mechanics. Section \ref{sec:experiment}
summarises the coupled-waveguide speed measurement. Section \ref{sec:objections}
addresses a set of typical objections: empirical equivalence, the
status of the measured speed, Rabi dynamics and non-separability,
wave-packet scattering, finite photon lifetime and pump-pulse duration,
and conditional versus unconditional dwell times. Section \ref{sec:bidirectional}
then presents a bidirectional Bohmian model as a proof of principle
that the reported discrepancy is not forced on trajectory theories
in general, but is tied specifically to the standard guiding equation.

\section{From density evolution to particle motion}

\label{sec:background}

In quantum mechanics, a freely moving wavepacket is the paradigmatic
example for defining a particle velocity. Such a wavepacket is naturally
associated with a moving particle, and its motion is characterised
by the group velocity, which is then identified with the particle's
velocity. More precisely, this assignment can be justified by examining
the particle density $\rho(\mathbf{x},t)=|\psi(\mathbf{x},t)|^{2}$,
which approximately moves according to 
\begin{equation}
\rho(\mathbf{x},t)\approx\rho(\mathbf{x}-\mathbf{v}t,0)\,.
\end{equation}
For a freely propagating, narrow-band wavepacket, this relation holds
approximately as long as dispersive broadening can be neglected. Crucially,
the example of a moving wavepacket illustrates that, in standard quantum
mechanics, the notion of particle velocity does not have the trajectory-level
meaning it has in Bohmian mechanics; it is tied to the behaviour of
statistical quantities, in particular the particle density distribution.

The situation is different in Bohmian mechanics. In this framework,
particles are assumed to have well-defined positions $\mathbf{x}(t)$
at all times, while the wavefunction $\psi(\mathbf{x},t)$ evolves
according to the Schrödinger equation. The central postulate is the
guiding equation, which prescribes the particle velocity as the ratio
of the quantum probability current $\mathbf{j}$ to the density $\rho$:
\begin{equation}
\frac{d\mathbf{x}}{dt}=\mathbf{v}_{S}(\mathbf{x},t)=\frac{\mathbf{j}(\mathbf{x},t)}{\rho(\mathbf{x},t)},\label{eq:guiding_equation}
\end{equation}
with 
\begin{equation}
\mathbf{j}(\mathbf{x},t)=\frac{\hbar}{m}\,\mathrm{Im}\!\left[\psi^{*}(\mathbf{x},t)\nabla\psi(\mathbf{x},t)\right].
\end{equation}
If initial particle positions are distributed according to the Born
rule, the ensemble of Bohmian trajectories reproduces the particle
density of standard quantum statistics.

The connection between $\rho$ and $\mathbf{j}$ is provided by the
continuity equation, 
\begin{equation}
\partial_{t}\rho(\mathbf{x},t)+\nabla\cdot\mathbf{j}(\mathbf{x},t)=0,
\end{equation}
which expresses local probability conservation. However, this connection
is not unique. For example, any modified flux of the form $\mathbf{j}'=\mathbf{j}+\mathbf{k}$
with $\nabla\cdot\mathbf{k}=0$ also satisfies the continuity equation,
leading to the same density evolution but to a different set of trajectories
\citep{de98,st08}. Another family of Bohmian models can be obtained
by decomposing the density and flux into components, $\rho(\mathbf{x},t)=\sum_{i}\rho^{(i)}(\mathbf{x},t)$
and $\mathbf{j}(\mathbf{x},t)=\sum_{i}\mathbf{j}^{(i)}(\mathbf{x},t)$,
with each component governed by its own guidance law. Such a decomposition
will be used in the discussion of the bidirectional Bohmian model.
The number of such constructions is unlimited. The standard guiding
equation does not follow from the Schrödinger equation or the continuity
equation alone, but requires additional assumptions. These assumptions
are precisely what our experiment puts under pressure: in the evanescent
regime, the standard Bohmian guiding equation assigns a state of motion
that does not agree with a physically plausible speed measurement.

\section{Particle speed in coupled waveguides}

\label{sec:experiment}

\begin{figure}
\centering \includegraphics[width=1\columnwidth]{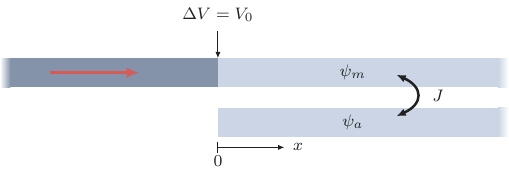} \caption{\textbf{Coupled waveguides.} A stream of particles (red arrow) is
confined in a waveguide potential and propagates towards a potential
step (vertical black arrow). At the step, a second waveguide opens
up. The particle transfer between the upper and the lower waveguides,
described by the coupling constant $J>0$, effectively acts as a clock
allowing particle speeds to be determined by considering the population
build-up in $\psi_{a}$.}
\label{fig:waveguides} 
\end{figure}

In our experiment \citep{sh25,kl23} we study the propagation of particles
in a coupled waveguide system. More specifically, we consider a system
where a stream of particles with mass $m$ is confined in a waveguide
potential and propagates towards a potential step at $x=0$ with $V(x)=0$
for $x<0$ and $V(x)=V_{0}$ for $x\ge0$. At the position of the
potential step, another waveguide potential opens up, which runs parallel
to the first one, see Fig.~\ref{fig:waveguides}. The potential barrier
between both waveguides is small enough that coupling between the
wave functions in the main ($\psi_{m}$) and auxiliary ($\psi_{a}$)
waveguide takes place and is quantitatively described by the coupling
constant $J>0$. For $x\ge0$, the steady state of the system is described
by coupled time-independent Schrödinger equations 
\begin{eqnarray}
E\psi_{m} & = & -\frac{\hbar^{2}}{2m}\frac{\partial^{2}\psi_{m}}{\partial x^{2}}+V_{0}\psi_{m}+\hbar J\left(\psi_{a}-\psi_{m}\right)\label{schroedinger_up}\\
E\psi_{a} & = & -\frac{\hbar^{2}}{2m}\frac{\partial^{2}\psi_{a}}{\partial x^{2}}+V_{0}\psi_{a}+\hbar J\left(\psi_{m}-\psi_{a}\right)\,.\label{schroedinger_down}
\end{eqnarray}
The total energy $E>0$ corresponds to the kinetic energy of the particles
before they hit the potential step. Moreover, we define the energy
difference $\Delta=E+\hbar J-V_{0}$, which plays the role of a local
kinetic-energy parameter in the step potential: like $T(x)=E-V(x)$,
it can be negative in the evanescent regime. The solution of this
system is 
\begin{eqnarray}
\psi_{m} & = & \frac{2k_{0}}{k_{0}+k_{2}}\cos(k_{1}x)\,\mathrm{e}^{ik_{2}x}\label{psi_up}\\
\psi_{a} & = & -\frac{2ik_{0}}{k_{0}+k_{2}}\sin(k_{1}x)\,\mathrm{e}^{ik_{2}x}\;,\label{psi_down}
\end{eqnarray}
with wavenumbers $k_{0,1,2}$ given by 
\begin{eqnarray}
k_{0} & = & \sqrt{2mE}/\hbar\label{k0}\\
k_{1} & = & \frac{mJ}{\hbar k_{2}}\label{k1}\\
k_{2} & = & \frac{\sqrt{2m}}{2\hbar}\left(\sqrt{\Delta+\hbar J}+\sqrt{\Delta-\hbar J}\right)\,.\label{k2}
\end{eqnarray}
In Eq.~\eqref{k2}, $\sqrt{x}$ denotes the positive real root for
$x>0$ and $i\sqrt{|x|}$ for $x<0$. 

The coupling constant governs the exchange of population between the
two waveguides: it controls how particles are distributed across the
guides, but not how they propagate along the waveguide axis (for $|\Delta|\gg\hbar J$,
see \citep{kl23}). In our analysis, we focus on a quantity that is
directly determined by the coupling, namely the relative population
of the two waveguides: 
\begin{align}
p_{a}(x) & =\frac{|\psi_{a}(x)|^{2}}{|\psi_{m}(x)|^{2}+|\psi_{a}(x)|^{2}}\label{eq:relativepopulation}\\
 & =\frac{|\sin(k_{1}x)|^{2}}{|\cos(k_{1}x)|^{2}+|\sin(k_{1}x)|^{2}}\,.
\end{align}
The common longitudinal factor in Eqs.~\eqref{psi_up} and \eqref{psi_down}
cancels from this normalised quantity: it is a common phase factor
in the propagating regime and a common attenuation factor in the evanescent
regime.

Physically, $p_{a}(x)$ characterises how far the particles present
at position $x$ have redistributed between the channels; it does
not specify what fraction of the incident particles reaches that position.
Propagating motion carries particles through successive positions
along the guides, whereas the evanescent solution describes a reflection
process, so not every particle that enters the region reaches a given
depth.

At short distances, the exact population response is 
\begin{equation}
p_{a}(x)=|k_{1}|^{2}x^{2}+\mathcal{O}\!\left((|k_{1}|x)^{4}\right).\label{eq:quadraticpopulation}
\end{equation}
For finite coupling, it therefore defines the inferred speed 
\begin{equation}
v_{J}=\frac{J}{|k_{1}|}=\frac{\hbar|k_{2}|}{m}.\label{eq:finite_coupling_speed}
\end{equation}
Around $\Delta=0$, its dependence on $J$ is appreciable and reflects
measurement back-action. For $|\Delta|\gg\hbar J$, Eq.~\eqref{k1}
gives $|k_{1}|\approx\sqrt{m/(2|\Delta|)}\,J$, so that 
\begin{equation}
p_{a}\approx\frac{f^{2}(Jx/v)}{g^{2}(Jx/v)+f^{2}(Jx/v)}\label{eq:relative_population_2}
\end{equation}
with $f(x)=\sin(x)$ and $g(x)=\cos(x)$ for $\Delta>0$, and $f(x)=\sinh(x)$
and $g(x)=\cosh(x)$ for $\Delta<0$. Here, we introduce the parameter
$v$, which is given by 
\begin{equation}
v=\sqrt{2|\Delta|/m}\,.\label{eq:speed}
\end{equation}
Note that $v\geq0$; hence, $v$ is a speed rather than a velocity.
Within each regime, $v$ and $J$ determine the relative population
profile in this weak-coupling limit. In particular, the speed $v$
governs how rapidly the population transfer builds up as a function
of the propagation distance for a given coupling $J$. The same expression
applies in both energy regimes, with trigonometric functions in the
propagating regime and hyperbolic functions in the evanescent regime.
A complementary formulation uses a local transition susceptibility,
which quantifies the response of a quantum state to weak coupling
into an orthogonal state \citep{KlaersSusceptibility26}. For a single
passage, the speed is the inverse of the suitably normalised susceptibility:
faster motion leads to less state conversion over a given interaction
length.

Experimentally, the speed $v$ can be inferred from the population
growth in the auxiliary waveguide, which is expected to increase quadratically
for small $x$: 
\begin{equation}
p_{a}(x)\approx\left(\frac{Jx}{v}\right)^{2}\,.\label{eq:speed_measure}
\end{equation}
By measuring $p_{a}(x)$ as a function of the propagation length and
comparing it with the predicted quadratic dependence, the ratio $J/v$
can be extracted. With an independently calibrated coupling $J$,
this directly determines $v$ without assuming a trajectory picture.
The exact finite-coupling expression and the size of its corrections
are discussed in Sec.~\ref{sec:clock_objection}.

We now turn to the main experimental observations \citep{sh25}. For
the speed inferred from the population transfer, we find a symmetric
energy--speed relation $v=\sqrt{2\left|\Delta\right|/m}$, where
$\Delta$ is the local kinetic-energy parameter in the step region
and changes sign between propagating and evanescent solutions. Thus,
the measured speed remains finite and follows the same functional
form in both regimes. The Bohmian velocity $v_{S}$ is measured through
an interference experiment. In the propagating regime, $v$ and $v_{S}$
are found to coincide. In the evanescent regime, however, $v_{S}$
becomes negligibly small and thus differs substantially from $v$.
From the perspective of Bohmian mechanics, the particles are therefore
essentially at rest, while our measurement indicates motion with several
thousand kilometres per second. This also has consequences for other
quantities, such as conditional and unconditional dwell times, which
are discussed in Sec.~\ref{sec:dwell_objection}.

\section{Objections and replies}

\label{sec:objections}

We now collect the main objections that have been raised against this
interpretation. The emphasis is on the conceptual structure of the
objections rather than on the specific wording of individual objections.

\subsection{Objection 1: Bohmian mechanics is empirically equivalent to quantum
mechanics}

A common objection is that our experiment cannot contradict Bohmian
mechanics unless it also conflicts with standard quantum theory. The
reason usually given is that Bohmian mechanics reproduces the same
wave functions and Born distributions as standard quantum mechanics,
and therefore reproduces the same particle densities and all quantities
derived from those densities \citep{Dr26,Di25,Wa25,Waegell26,DM26,Daem26,DrNabet26,Beck26}.

What is established by the usual equivariance argument is that Bohmian
mechanics reproduces the same particle density as standard quantum
mechanics, provided the initial distribution is the Born distribution.
We do not question this. The failure to reproduce the density has
never been part of our argument. The discrepancy is more specific:
the outcome of a physically plausible speed measurement does not agree
with the motional state of the particles assumed in the standard formulation
of Bohmian mechanics. This is possible because density reproduction
alone does not uniquely fix the guiding law within Bohmian-type theories.
The standard guiding equation is therefore an additional physical
postulate, not a consequence of empirical equivalence alone; it can
be scrutinised if the motional state it assigns conflicts with an
independently motivated speed measurement.

\subsection{Objection 2: comparison of different quantities}

A second objection states that the speed inferred from the relative
population density in the two waveguides is simply a different quantity
from the Bohmian velocity \citep{Di25,Wa25,Dr26,Waegell26,YeMeasured25,YeActual26,ShangInstant25,DM26,Beck26}.
In one sense, this is correct. The speed measured in our experiment
is not obtained from the phase gradient, or equivalently from the
current-to-density ratio used in the standard guiding equation, Eq.
\eqref{eq:guiding_equation}. However, this is not a drawback, but
a prerequisite for an independent check.

From the perspective of standard Bohmian mechanics, particle motion
is described by the velocity obtained from the phase gradient of the
wavefunction, as specified by the standard guiding equation, Eq. \eqref{eq:guiding_equation}.
This quantity is measurable, and we determine it in our experiment
using an interference measurement. However, interpreting the phase-gradient
measurement as a velocity measurement already assumes that the standard
guiding equation correctly captures the motion of the particles. That
is precisely the point being tested. As noted above, the standard
guiding equation does not follow uniquely from the continuity equation;
additional assumptions are required. Testing such an assignment therefore
requires a notion of velocity or speed obtained by different physical
means, without presupposing the very equation under examination.

In the propagating regime, this check succeeds: the speed extracted
from the population transfer in the waveguides agrees with the standard
Bohmian velocity obtained from the phase gradient. The evanescent
regime, however, is different. There, our results for the speed of
the particles disagree with the Bohmian velocity. This disagreement
can only be removed by providing a non-circular argument showing why
the speed inferred from the coupled waveguide system, which gives
the expected result in the propagating regime, ceases to provide a
meaningful speed measurement in the evanescent regime. Such an argument
cannot simply use the standard Bohmian velocity in that regime as
the criterion, since that is exactly the assignment being tested.
Otherwise the disagreement is resolved only by assuming the conclusion.
Proposed criteria based on Rabi dynamics and separability are examined
in Sec.~\ref{sec:clock_objection}.

\subsection{Objection 3: the absence of Rabi oscillations and non-separability
invalidate the transverse coupling as a clock in the evanescent regime}

\label{sec:clock_objection}

Beck et al. argue that the speed inference assumes autonomous double-well
dynamics with a tunnelling time set by the transverse coupling. They
contend that longitudinal--transverse entanglement disrupts these
dynamics in the evanescent regime \citep{Beck26}.

Two issues must be separated: Rabi dynamics and non-separability.
First, our measurement scheme does not rely on Rabi oscillations in
either regime. As shown below, their presence or absence does not
enter our speed inference.

The short-distance population response is given in Eq.~\eqref{eq:quadraticpopulation}.
The common longitudinal evolution cancels from the normalised quantity
in Eq.~\eqref{eq:relativepopulation}. In the fully propagating regime
($\Delta>\hbar J$), $k_{1}$ is real; in the fully evanescent regime
($\Delta<-\hbar J$), it is purely imaginary. Writing $\tilde{x}=|k_{1}|x$,
the corresponding responses expand as 
\begin{equation}
\begin{aligned}p_{a}^{\mathrm{prop}}(x) & =\tilde{x}^{2}-\frac{1}{3}\tilde{x}^{4}+\mathcal{O}(\tilde{x}^{6}),\\
p_{a}^{\mathrm{ev}}(x) & =\tilde{x}^{2}-\frac{5}{3}\tilde{x}^{4}+\mathcal{O}(\tilde{x}^{6}).
\end{aligned}
\label{eq:populationexpansions}
\end{equation}
Thus the quadratic response used by the measurement is identical in
the propagating and evanescent regimes. Only at the next order do
the two responses differ. These differences at larger distances, including
the presence or absence of Rabi oscillations, do not enter the speed
inference in our experiment.

Second, separability means that the full wave function factorises
as $\psi(x,y)=\alpha(x)\beta(y)$. The operational consequence of
non-separability in the context of our experiment is a possible dependence
of the inferred result on the coupling $J$. This residual dependence
quantifies the measurement back-action associated with the non-separability
of the wavefunction. It would compromise the measurement scheme if
measurement back-action were substantially different in the propagating
and evanescent cases. However, the exact solution of the coupled-waveguide
system shows no such asymmetry.

Equations~\eqref{k2} and \eqref{eq:finite_coupling_speed} give
the exact result \citep{kl23} 
\begin{equation}
v_{J}(\Delta)=\begin{cases}
\sqrt{\hbar J/m}, & |\Delta|\leq\hbar J,\\[3pt]
{\displaystyle \sqrt{\frac{|\Delta|+\sqrt{\Delta^{2}-(\hbar J)^{2}}}{m}},} & |\Delta|>\hbar J.
\end{cases}\label{eq:generalvelocity}
\end{equation}
It depends only on $|\Delta|$, so the propagating and evanescent
branches have exactly the same dependence on the coupling. Around
$|\Delta|\sim\hbar J$, the inferred speed depends appreciably on
$J$, indicating significant measurement back-action. For $|\Delta|\gg\hbar J$,
a Taylor expansion gives 
\begin{equation}
\begin{aligned}v_{J}(\Delta)=\sqrt{\frac{2|\Delta|}{m}}\bigg[1 & -\frac{1}{8}\left(\frac{\hbar J}{|\Delta|}\right)^{2}\\
 & +\mathcal{O}\!\left(\left(\frac{\hbar J}{|\Delta|}\right)^{4}\right)\bigg].
\end{aligned}
\label{eq:velocityexpansion}
\end{equation}
The leading $J$-dependent correction is therefore quadratic, and
the same $J$-independent limit in Eq.~\eqref{eq:speed} is reached
from the propagating and evanescent sides. The relevant crossover
is $|\Delta|\sim\hbar J$, not the boundary between propagating and
evanescent motion. At finite $J$, the channel and longitudinal degrees
of freedom are non-separable in both regimes; the expansion above
quantifies its operational effect and shows that it becomes equally
weak on both sides.

Taken together, Eqs.~\eqref{eq:quadraticpopulation} and \eqref{eq:populationexpansions}--\eqref{eq:velocityexpansion}
give a direct result: the propagating--evanescent asymmetry sought
by Beck et al. is absent from the measurement protocol. It appears
neither in the leading short-distance response used to infer the speed
nor in the coupling-dependent measurement back-action.

Any stronger conclusion that the evanescent result cannot represent
a plausible particle speed \citep{Beck26} is also contradicted by
a concrete construction. In Sec.~\ref{sec:bidirectional}, we introduce
a bidirectional Bohmian model that reproduces the standard stationary
density and its vanishing net current while assigning finite right-
and left-moving particle velocities in the evanescent region. Their
speed is compatible with the speed inferred by our waveguide measurement.
This model provides an explicit trajectory interpretation of the measured
evanescent speed within a Bohmian framework. The disagreement is therefore
specific to the standard guiding law, not to the possibility of interpreting
the measured speed as particle motion.

\subsection{Objection 4: the experiment should be described by wave-packet scattering}

\label{sec:wavepacket_objection}

Several responses argue that a stationary treatment is insufficient
for describing particle motion in the evanescent regime, and that
the formation of the observed density must instead be analysed through
time-dependent wave-packet scattering or transient dynamics \citep{Di25,Dr25,Dr26,Daem26,Waegell26,ShangInstant25,DM26}.
In this scenario, the phase gradient need not vanish at the potential
barrier, and the transmitted component can enter the forbidden region
with a non-zero Bohmian velocity field.

This objection is based on a misunderstanding of the physical situation
realised in our experiment \citep{sh25}. Our system operates in a
regime that is well described by single-mode lasing. Under non-resonant
optical pumping, spontaneously emitted photons randomly populate the
eigenmodes of the system. Gain competition then selects a single dominant
mode, which is subsequently macroscopically amplified. This corresponds
to a single energy eigenmode of the system. As a result, the particle
density as a function of space and time is, to a very good approximation,
given by 
\begin{equation}
\rho(\mathbf{x},t)=N(t)|\phi(\mathbf{x})|^{2},
\end{equation}
where $N(t)$ is a time-dependent global intensity, or density, and
$\phi(\mathbf{x})$ is the normalised spatial profile of the selected
energy eigenmode. In this sense, the system realises an energy eigenstate
to a very good approximation, with temporal dynamics restricted to
an overall scaling factor.

The finite pump duration controls the global factor $N(t)$; it is
not the temporal width of a longitudinal scattering packet and therefore
does not determine the packet's wave-number width $\sigma_{k}$ \citep{Beck26}.
The spatial scales already illustrate the problem with identifying
these two durations. A longitudinal packet with the pump-pulse duration
of $26\,\mathrm{ns}$ and a representative propagating speed of $2000\,\mathrm{km\,s^{-1}}$
would have a spatial extent of about $5\,\mathrm{cm}$. This is much
larger than the entire waveguide structure, which extends from $x=-900\,\mu\mathrm{m}$
to $x=500\,\mu\mathrm{m}$ \citep{sh25}. Such a pulse-length envelope
does not represent a localised packet entering and leaving the step
on the scale of the apparatus. Its duration cannot simply be used
to set the spectral width of the transient scattering model.

\subsection{Objection 5: non-idealities remove the discrepancy}

A further objection is that the real cavity is not perfectly ideal
\citep{Di25,Dr25,Dr26}, and that the associated corrections invalidate
the reported challenge to Bohmian mechanics \citep{DrNabet26}. The
dominant non-ideality is the finite photon lifetime, associated with
radiative leakage; the finite duration of the pump pulse is secondary.
Both effects can be included in more detailed models, but they are
too small to have a significant impact on the experiment. In particular,
the measured decay constants shown in Fig.~2c of our Nature article
\citep{sh25} are reproduced by the loss-free model within the relevant
experimental accuracy.

The finite lifetime does not spoil the eigenmode picture. In a standard
loss model, the resonance energy, or equivalently the effective potential,
acquires a small imaginary part. The corresponding in-plane wave vectors
are then slightly complex, which introduces a small phase gradient
and hence a small non-zero Bohmian velocity. This is a refinement
of the quasi-stationary single-mode description, not a replacement
of it by a different scattering problem. Quantitatively, Drezet et
al. estimate the leakage-induced longitudinal Bohmian velocity to
be of order $30\,\mathrm{km\,s^{-1}}$, while the waveguide-inferred
speed scale they compare to is of order $2000\,\mathrm{km\,s^{-1}}$
\citep{Dr25,Dr26}. They also note that the correction associated
with the finite duration of the pump pulse is about two orders of
magnitude smaller than the leakage correction. Thus the dominant non-ideality
can make the ideal Bohmian velocity strictly non-zero, but it does
not bridge the gap to the measured speed.

This is the sense in which we speak of a vanishing Bohmian velocity
in our paper: the directly measured Bohmian velocity is consistent
with zero within the experimental uncertainty. The non-ideal corrections
discussed by Drezet et al. leave intact the central disagreement between
the speed inferred from the coupled waveguides and the velocity assigned
by the standard guiding equation.

\subsection{Objection 6: the dwell-time comparison mixes conditional and unconditional
quantities}

\label{sec:dwell_objection}

Related criticisms of the dwell-time comparison concern the definition
of the incident current and its decomposition into incoming and reflected
contributions~\mbox{\citep{Di25,YeMeasured25,DM26}}.

Beck et al. argue that the unconditional dwell time is an average
over all incident particles, including those reflected before entering
the step, whereas an estimate based on the amplitude decay length
and speed concerns particles that enter. They conclude that the dwell
time must be conditioned on entry before the two quantities can be
compared \citep{Beck26}. We agree with this conceptual distinction.
In the Nature paper, the comparison was intended as a consistency
check at the level of an estimate, rather than an assertion that the
conditional and unconditional times are identical \citep{sh25}.

For a single potential step, the relation between the two quantities
can be calculated exactly. We now take $J=0$ and $0<E<V_{0}$, with
$V(x)=0$ for $x<0$ and $V(x)=V_{0}$ for $x\geq0$. The stationary
scattering solution is 
\begin{equation}
\psi(x)=\begin{cases}
e^{ikx}+r\,e^{-ikx}, & x<0,\\[4pt]
t\,e^{-\kappa x}, & x>0,
\end{cases}\label{eq:wavefunction_step}
\end{equation}
where $k=\sqrt{2mE}/\hbar$ and $\kappa=\sqrt{2m(V_{0}-E)}/\hbar$.
The reflection and evanescent amplitudes are 
\begin{equation}
r=\frac{k-i\kappa}{k+i\kappa},\qquad t=\frac{2k}{k+i\kappa}.\label{eq:step_amplitudes}
\end{equation}
The incident flux is $j_{\mathrm{in}}=\hbar k/m$, and the density
is 
\begin{equation}
\rho(x)=|\psi(x)|^{2}=\begin{cases}
|e^{ikx}+re^{-ikx}|^{2}, & x<0,\\[4pt]
|t|^{2}e^{-2\kappa x}, & x>0.
\end{cases}\label{eq:density_step}
\end{equation}
The unconditional dwell time is defined by the integrated density
divided by the incident flux \citep{ha89,la94,le90}: 
\begin{equation}
\begin{aligned}\tau_{\mathrm{dwell}}^{\mathrm{QM}} & =\frac{1}{j_{\mathrm{in}}}\int_{0}^{\infty}\rho(x)\,dx=\frac{|t|^{2}}{2\kappa j_{\mathrm{in}}}\\
 & =\frac{2mk}{\hbar\kappa(k^{2}+\kappa^{2})}.
\end{aligned}
\label{eq:standard_dwell}
\end{equation}
On the other hand, with the amplitude decay length $\lambda=1/\kappa$
and the evanescent speed $v=\hbar\kappa/m$, the conditional residence-time
estimate is 
\begin{equation}
\tau_{\lambda}=\frac{\lambda}{v}=\frac{m}{\hbar\kappa^{2}}.\label{eq:conditional_estimate}
\end{equation}
Here $\lambda$ is the amplitude decay length; the density decay length
is $1/(2\kappa)$. The exact correction factor is therefore 
\begin{equation}
\begin{aligned}F=\frac{\tau_{\lambda}}{\tau_{\mathrm{dwell}}^{\mathrm{QM}}} & =\frac{k^{2}+\kappa^{2}}{2k\kappa}\\
 & =\frac{V_{0}}{2\sqrt{|\Delta|(V_{0}-|\Delta|)}},
\end{aligned}
\label{eq:dwell_factor}
\end{equation}
with $\Delta=E-V_{0}<0$ in this uncoupled problem.

\begin{figure}
\centering \includegraphics[width=1\columnwidth]{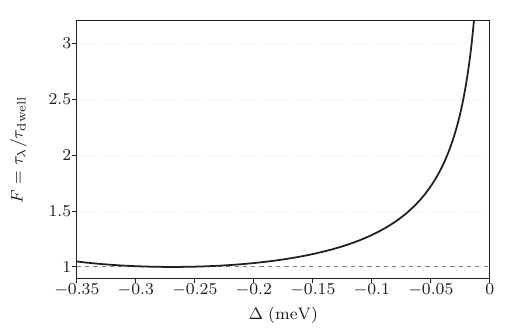}
\caption{\textbf{Conditional residence-time estimate and unconditional dwell
time.} Exact correction factor $F$ in Eq.~\eqref{eq:dwell_factor}
for a single potential step with $V_{0}=0.538\,\mathrm{meV}$, over
the energy range of Fig.~3b of Ref.~\citep{sh25}. The dashed line
marks $F=1$. The two times are close over a broad part of this range;
their difference increases towards the threshold $\Delta=0$.}
\label{fig:dwell_factor} 
\end{figure}

The factor equals one at $k=\kappa$ and remains close to unity over
a broad part of the experimentally relevant parameter range (Fig.~\ref{fig:dwell_factor}).
This explains quantitatively why $\lambda/v$ provides a useful estimate
of the unconditional dwell time over that range in Fig.~3b of the
Nature paper. The departure towards $\Delta=0$ is also described
by the same factor; equality is not claimed across the full range.
The distinction between the two quantities therefore does not invalidate
the dwell-time consistency check. Section~\ref{sec:bidirectional}
gives the corresponding trajectory account, including the entry probability
that relates the conditional residence time to the unconditional dwell
time.

\section{A bidirectional Bohmian model}

\label{sec:bidirectional}

The preceding sections raise a natural question about the disagreement
between the speed inferred from the coupled waveguides and the velocity
assigned by the standard Bohmian guiding equation: Is it a necessary
consequence of introducing particle trajectories, or is it tied to
this particular guiding law? The following model is meant to address
this question in a limited and constructive way. It is not proposed
as a replacement for the full Bohmian theory, nor as a complete ontology
for all quantum systems. Rather, it is a minimal example showing how
a trajectory model can reproduce the same stationary density while
assigning finite speeds in the evanescent region. In this model, we
introduce right- and left-moving subensembles. The standard probability
current retains its usual meaning: it is the net current obtained
after adding the right- and left-moving contributions. For the evanescent
region this net current vanishes, even though the two directional
subensembles move locally with finite speed. 
\begin{figure}
\begin{centering}
\includegraphics[width=1\columnwidth]{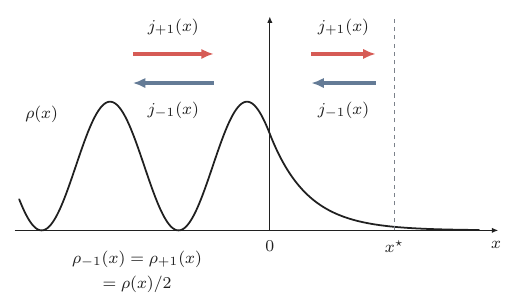} 
\par\end{centering}
\caption{\textbf{Bidirectional Bohmian model at a potential step.}{} The stationary
density is the usual quantum density: an interference pattern for
$x<0$ and an exponential tail for $x>0$. The model decomposes the
ensemble into right- and left-moving components with currents $j_{+1}$
and $j_{-1}$. Their sum gives the standard net current. In the given
scenario this net current vanishes, $j_{+1}+j_{-1}=0$, although each
component is finite. In the evanescent region, right-movers propagate
up to a randomly sampled return point $x^{\star}$, after which they
move back to the left. }
\label{fig:bidirectional_model} 
\end{figure}

The model extends the particle position by a binary variable indicating
the direction of motion. Thus the hidden variables are the position
and the direction label, rather than the position alone. The continuous
motion follows deterministic guiding rules, with stochastic sampling
confined to entry decisions and turning points. The bidirectional
construction discussed in this section is shown schematically in Fig.~\ref{fig:bidirectional_model}.
We use the loss-free, one-dimensional step problem ($J=0$) introduced
in Sec.~\ref{sec:dwell_objection}, with the stationary state and
density given by Eqs.~\eqref{eq:wavefunction_step}--\eqref{eq:density_step}.

In this model, each particle carries an additional discrete label
$\sigma=\pm1$ that specifies the direction of motion: $\sigma=+1$
corresponds to a particle moving to the right, and $\sigma=-1$ to
a particle moving to the left. The two directional subensembles add
up to the quantum density $\rho(x)$, while their currents add up
to the net flux. The guiding rules are defined as follows. For $x>0$,
which we refer to as the \emph{right} or \textit{evanescent} region,
particles move with constant speed $v=\sqrt{2(V_{0}-E)/m}$ according
to the guiding equation $\dot{x}=\sigma v$, where $\sigma=\pm1$
indicates the direction of motion. Note that the speed $v$ corresponds
to the speed in Eq.~\eqref{eq:speed}. Each right-mover ($\sigma=+1$)
is assigned a random turning point $x^{\star}>0$, drawn from the
exponential distribution $p(x^{\star})=2\kappa e^{-2\kappa x^{\star}}$.
At this point it flips to $\sigma=-1$ and propagates back to the
left with velocity $-v$. For $x<0$, which we refer to as the \textit{left}
region, we take the particle fluxes associated with the two directions
to be constant, equal in magnitude but opposite in sign, $j_{+1}(x)=-j_{-1}(x)=j_{\mathrm{in}}$.
Thus, the net flux vanishes. The guiding equation is $\dot{x}=\sigma\,2j_{\mathrm{in}}/\rho(x)$.

Boundary rules at $x=0$ are defined as follows. When a right-moving
particle in the left region reaches the boundary, it undergoes one
of two possible outcomes: with probability $a$, it enters the evanescent
region as a right-mover with a freshly sampled turning point $x^{\star}$;
with probability $1-a$, it is immediately reflected and turned into
a left-mover ($\sigma=-1$) at $x=0$. Left-movers originating from
the right region cross the boundary at $x=0$ without reflection and
continue their motion in the left region as left-movers.

To fix the parameter $a$, we notice that the incident plane wave
is normalised to unit amplitude (see Eq.~\eqref{eq:wavefunction_step}).
This sets the incident flux to $j_{\mathrm{in}}=\hbar k/m$. The flux
of right-movers injected into the evanescent region is then $a\,j_{\mathrm{in}}$.
Since right-movers in $x>0$ move with speed $v$, their stationary
contribution to the density is $(a\,j_{\mathrm{in}}/v)\,e^{-2\kappa x}$.
The returning left-movers contribute the same amount, so the total
density in the evanescent region is $\rho(x)=2(a\,j_{\mathrm{in}}/v)\,e^{-2\kappa x}$.
Matching this to the quantum result Eq.~\eqref{eq:density_step}
gives 
\begin{equation}
a=\frac{|t|^{2}v}{2j_{\mathrm{in}}}\,.\label{eq:entry_probability}
\end{equation}
Indeed, this gives $a=2k\kappa/(k^{2}+\kappa^{2})$, and therefore
$0\le a\le1$. Comparison with Eq.~\eqref{eq:dwell_factor} gives
$a=1/F$.

As an application, the model provides a trajectory interpretation
of the dwell-time relation in Sec.~\ref{sec:dwell_objection}. Particles
enter the right region with probability $a$, move at constant speed
$v$ up to a random turning point with $\langle x^{\star}\rangle=1/(2\kappa)$,
and return, so that the mean residence time conditional on entry is
$1/(\kappa v)=\tau_{\lambda}$. The mean round-trip path is $2\langle x^{\star}\rangle=1/\kappa=\lambda$,
which accounts for the decay-length convention in Eq.~\eqref{eq:conditional_estimate}.
Per incident particle, including immediate reflections with zero residence
time, this gives 
\begin{equation}
\tau_{\mathrm{dwell}}^{\mathrm{BBM}}=\frac{a}{\kappa v}=a\,\tau_{\lambda}=\tau_{\mathrm{dwell}}^{\mathrm{QM}}.\label{eq:model_dwell}
\end{equation}
Thus the model reproduces the standard unconditional dwell time exactly,
while retaining the distinction from the conditional residence time.
If the same evanescent-speed assignment is applied to a finite opaque
barrier of width $L$, the corresponding tunnelling time is obtained
by dividing the barrier width by the evanescent speed, $\tau_{\mathrm{BL}}=L/v=L\sqrt{m/(2(V_{0}-E))}$,
which is the Büttiker--Landauer tunnelling time \citep{bu82,la94}.

\section{Summary}

Our experiment \citep{sh25} infers particle speed from a comparison
of longitudinal motion and transverse population transfer in a coupled-waveguide
system. The inferred speed agrees with the velocity assigned by the
standard Bohmian guiding equation in the propagating regime but disagrees
with it in the evanescent regime.

The objections discussed above do not remove this point. The short-distance
population response used for the measurement does not require autonomous
Rabi oscillations, and the finite-coupling correction has the same
form on the propagating and evanescent sides. Conditional residence
times and the unconditional dwell time are distinct, but their exact
relation is consistent with the original comparison as an estimate.
Wave-packet scattering describes a different physical regime, and
non-ideal corrections from finite photon lifetime and pump-pulse duration
are too small to account for the discrepancy between the speed inferred
from the coupled waveguides and the velocity assigned by the standard
guiding equation.

The bidirectional Bohmian model shows that the disagreement is not
a consequence of introducing trajectories as such, since a trajectory
model can reproduce the same stationary density while assigning finite
speeds in the evanescent region. Agreement or disagreement with the
Bohmian framework therefore depends on the choice of the guiding equation
that is assumed. Our analysis of the objections therefore reaffirms
the challenge posed by our experiment to the standard Bohmian guiding
equation as a general law of particle motion.

\bibliography{references}

\end{document}